\documentclass[twocolumn]{aastex631}
\usepackage{graphicx}
\usepackage{bm}
\usepackage{physics}
\usepackage{booktabs} 
\usepackage{makecell}
\usepackage{threeparttable}
\usepackage{multirow}

\shorttitle{BSI in AGN disk}
\shortauthors{Wang et al.}
\graphicspath{{./}{figure/}}
\newcommand{\R}[1]{\textcolor{red}{#1}}
\newcommand{\B}[1]{\textcolor{blue}{#1}}
\usepackage{hyperref}
\usepackage{natbib}
\defcitealias{Wang2025arXiv}{Paper I}

\begin{document}
\title{Simulation of Binary–Single Interactions in AGN Disks II: Merger Probability of Binary Black Holes during Chaotic Triple Process}  

\author[0000-0001-5019-4729]{Mengye Wang}
\affiliation{Department of Astronomy, School of Physics, Huazhong University of Science and Technology, Luoyu Road 1037, Wuhan, China}

\author[0000-0003-4773-4987]{Qingwen Wu}
\altaffiliation{qwwu@hust.edu.cn} 
\affiliation{Department of Astronomy, School of Physics, Huazhong University of Science and Technology, Luoyu Road 1037, Wuhan, China}

\author[0000-0001-7192-4874]{Yiqiu Ma}
\altaffiliation{myqphy@hust.edu.cn}
\affiliation {National Gravitation Laboratory, School of Physics, Huazhong University of Science and Technology, Luoyu Road 1037, Wuhan, China}
\affiliation{Department of Astronomy, School of Physics, Huazhong University of Science and Technology, Luoyu Road 1037, Wuhan, China}





\begin{abstract}
Stellar-mass binary black hole\,(BBH) mergers resulting from binary–single interactions\,(BSIs) in active galactic nucleus\,(AGN) disks are a potential source of gravitational wave\,(GW) events with measurable eccentricities.  
Previous hydrodynamical simulations have shown that ambient gas can significantly influence the dynamics of BSIs. However, due to limitations such as the use of purely Newtonian dynamics and small sample sizes, a direct estimation of the BBH merger probability during BSI has remained elusive. In this work, we directly quantify the merger probability, based on a suite of 1800 two-dimensional hydrodynamical simulations coupled with post-Newtonian \emph{N}-body calculations.  
Our results demonstrate that dense gas can enhance the merger probability by both shrinking the spatial scale of the triple system and increasing the number of binary–single encounters. These two effects together boost the merger probability by a factor of $\sim$5, from 4\% to as high as 20\%.
Among the two effects, our analysis suggests that the increase in encounter frequency plays a slightly more significant role in driving the enhancement.
Moreover, this enhancement becomes more significant at larger radial distances from the central SMBH, since the total gas mass enclosed within the Hill sphere of the triple system increases with radius. Finally, the BSI process in AGN disks can naturally produce double GW merger events within a timescale of $\sim$year, which may serve as potential observational signatures of BSI occurring in AGN disk environments.

\end{abstract}


\section{Introduction}    \label{sec:intro}
Binary black hole\,(BBH) merger in active galactic nucleus\,(AGN) disk is a potential formation channel for gravitational wave\,(GW) events\,\citep[e.g.,][]{McKernan2014MNRAS,Bartos2017NatCo,stone2017MNRAS}. Due to in-situ formation and capture from the nuclear star cluster, AGN disk naturally host a large population of stellar-mass black holes\,\citep[sBHs, e.g.,][]{Bartos2017ApJ,Panamarev2018MNRAS,Yang2019ApJ,Pan2021PhRvD,Wang2023MN,LYP2024ApJ,LYP2025PhRvD}. Moreover, many hydrodynamical\,(HD) simulations suggested that when two sBHs undergo a close encounter in a gas-rich environment, their circumsingle disks\,(CSDs) will collide. The accumulated material between the two sBHs following the CSD collision leads to energy dissipation within the binary system, facilitating the formation of a bound BBH\,\citep[e.g.,][]{LJR2022ApJ,LJR2023ApJ,Rowan2023MNRAS,Rowan2024MNRAS,Whitehead2024MNRAS,Whitehead2024MNRASb,Whitehead2025arXiv}. These newly formed BBH systems, embedded within the AGN disk, may experience continuous orbital shrinkage\,\citep[e.g.,][]{Baruteau2011ApJ,LYP2021ApJ911,LYP2022ApJ,LRX2022MNRAS,Dempsey2022ApJ,Mishra2024arXiv,Dittmann2024ApJ,Dittmann2025arXiv} or runaway eccentricity growth\,\citep[][]{Calcino2024ApJ}, ultimately leading to their merger.  The merger of these BBHs, occurring in the gas-rich environment surrounding supermassive black holes\,(SMBHs), may provide a possible explanation for the presence of mass-gap black holes\,\citep[][]{Yang2019PhRvL,Tagawa2021ApJ}, non-zero effective spins\,\citep[][]{Cook2024arXiv}, and non-zero eccentricities\,\citep[e.g.,][]{Graham2020PhRvL} in GW events detected by ground-based detectors\,\citep[][]{Acernese2015,LIGO2015,Kagra2019NatAs}.

In addition to the influence of ambient gas, a BBH embedded in a gas-rich environment may undergo dynamical encounters with another single sBH, a process known as binary–single interaction\,(BSI; e.g., \citealt{Valtonen2006tbp,Samsing2014ApJ,Giant2021PhRvX,Giant2023MNRAS,Trani2024A&A}). During such chaotic three-body encounters, close passage between two sBHs can lead to rapid orbital energy loss via GW emission, potentially resulting in prompt mergers\,(e.g., \citealt{Samsing2018MNRAS, Rodriguez2018PhRvD}).

In globular clusters or nuclear stellar clusters, the merger probability during BSI is typically much lower than that of isolated BBHs. However, recent studies show that in coplanar AGN disks—even without gas effects—the merger probability during BSI can be enhanced by up to two orders of magnitude compared to spherical cluster environments\,\citep{Samsing2022Natur, Fabj2024MNRAS}. Compared to isolated BBH mergers, BSI-driven mergers exhibit distinct observational signatures, such as high orbital eccentricities\,\citep[e.g.,][]{Rodriguez2018PhRvD} and double GW mergers\,\citep[e.g.,][]{Samsing2018MNRAS,Samsing2019MNRASb}.

Intriguingly, reanalyses of GW events have yielded hints of these signatures: (1) \emph{Eccentric Mergers:} A reanalysis of 17 GW events by \citet{Planas2025arXiv} found statistically significant non-zero eccentricities in two BBH mergers\,\citep[with tentative hints in two others, see also][]{Gamba2023NatAa}. The detected eccentricities, measured at $10\,\mathrm{Hz}$, ranged from 0.2–0.3 for systems with total masses of $\sim100\,M_\odot$. Maintaining such high eccentricities at $10\,\mathrm{Hz}$ requires extreme initial eccentricities, most naturally produced in multi-body dynamical encounters\,\citep[see][]{Rodriguez2018PhRvD}; (2) \emph{Double GW Mergers:} Observed correlations between GW events indicate that the GW170809–GW151012 and GW190514–GW190521 pairs are potential candidates for double GW mergers scenario\,\citep[e.g.,][]{Veske2020MNRAS, Veske2021ApJ, Li-Fan2025arXiv, Li-Fan2025arXiv2}. In such a scenario, the remnant of a BBH merger subsequently merges with a third sBH—a signature strongly linked to BSI dynamics \citep{Samsing2018PhRvD,Samsing2019MNRASb, Liu2021MNRAS, Lu2021MNRAS, Gayathri2023arXiv}.

In our previous work (\citealt{Wang2025arXiv}; hereafter \citetalias{Wang2025arXiv}), we performed coupled HD and \emph{N}-body simulations to investigate the dynamical processes of BSI in AGN disks (see also the related work by \citealt{Rowan2025arXiv}). Our results demonstrated that the surrounding gas can efficiently reduce the triple system's spatial scale and increase the number of binary-single encounters during BSI, which may significantly enhance the BBH merger probability during BSI in AGN disks. However, we did not quantify the BBH merger probability due to two primary limitations: (1) the inability of Newtonian \emph{N}-body simulations to capture relativistic effects during the close encounter and the late inspiral phase of BBHs; (2) the relatively small sample size of our simulations.

To address the first issue, \cite{Rowan2025arXiv} estimated the enhancement of merger probability due to gas by employing the analytical approximation from \cite{Fabj2024MNRAS}, under the condition of only considering the effect of gas-induced shrinkage of the triple system's scale. They found that the spatial scale of the end-state triple system could be reduced by two to three orders of magnitude compared to the initial scale. In this way, the merger probability can be enhanced by a factor of $\sim 3.5-8$. This approach could be affected by the following factors that need to be further investigated.
(1) Replacing the triple system's scale throughout the entire process with its end-state scale likely overestimates the enhancement;
(2) It neglects the increase in the number of binary-single encounters caused by gas drag.
(3) In such a chaotic BSI process, the GW radiation between the sBHs and the gravitational interaction with gas are coupled, and separating them could affect the results. Thus, in this work, we incorporate post-Newtonian\,(PN) corrections into our \emph{N}-body simulations and perform coupled HD and \emph{N}-body simulations with a larger sample size to directly obtain the BBH merger probability during BSI in AGN disk. Furthermore, we compare our results with predictions from the gas dynamical friction\,(GDF) model.


The structure of this paper is as follows. Section\,\ref{sec:model} provides a brief overview of the numerical methods and introduces the initial parameter settings. In Section\,\ref{sec:results}, we present the BBH merger probability during BSI based on a suite of HD simulations, test its radial dependence on the AGN disk, and compare our HD results with the GDF model. In Section\,\ref{sec:discussion}, we discuss the importance of direct PN simulation, the conditions for BSI occurrence, the possible observational features of BSI in AGN disk, and some limitations of this work. Finally, we summarize our main results in Section\,\ref{sec:summary}.

\section{METHOD} \label{sec:model}
In this work, we perform two-dimensional simulations of the BSI process in the AGN disk using the Eulerian code Athena++\,\citep{stone2020ApJS} coupled with the $N$-body code REBOUND\,\citep{Rein2012A&A}, where we incorporate GW radiation through 2.5-orders PN formalism in the $N$-body integration.
Our primary system and parameter settings have been detailed in \citetalias{Wang2025arXiv}, and a brief overview of the model is presented here.

\subsection{Hydrodynamics and \emph{N}-body dynamics}
Our model consists of a central SMBH with mass\,\( M_\bullet = 10^8\,M_\odot \), its accretion disk, and three sBHs, initially configured as a BBH and a single sBH. The three sBHs are placed within a two-dimensional rectangular shearing box at radius \( R_0 \) on the AGN disk, which serves as the primary domain of our simulation. The characteristic length scale in the shearing box is defined by the Hill radius of the three sBHs,
\begin{equation}
    R_{\rm H} = R_0 \left(\frac{M_{\rm sBH}}{3M_\bullet}\right)^{1/3},
\end{equation}
where $M_{\rm sBH}=m_1+m_2+m_3$ is the total mass of the three sBHs. 

The gas dynamics is governed by the hydrodynamical equations:
\begin{equation}
\begin{split}
 & \frac{\partial \Sigma}{\partial t} + \nabla \cdot (\Sigma \bm{u}) = 0, \\
  & \frac{\partial (\Sigma \bm{u}) }{\partial t} + \nabla\cdot (\Sigma \bm{u}\bm{u}) + \nabla P =\\
  & \Sigma\left[2\bm{u}\times (\Omega_0\bm{\hat{z}})+ 2q\Omega^2_0 x\bm{\hat{x}} + \nabla \cdot \bm{T} -\nabla \Phi_{\rm sBH}\right],
\end{split}
\end{equation}
where $\Sigma$, $\bm{u}$ and $P$ are gas surface density distribution, velocity fields and the pressure, and $q \equiv -{\rm d\,ln}\, \Omega/ {\rm d\,ln}\, R=3/2$ is the background shear parameter of a Keplerian disk, $\Omega_0\equiv(GM/R_0^3)^{1/2}$ is the angular velocity of the shearing box orbiting around the SMBH. The $ \bm{\hat x}, \bm{\hat y}$ and $\bm{\hat z}$ are the unit vectors in the corotating frame.
$\bm{T}$ is the viscous stress tensor with an isotropic kinematic viscosity $\nu$:
\begin{equation}
    T_{ij} = \nu \left( \frac{\partial u_i}{\partial x_j} + \frac{\partial u_j}{\partial x_i} - \frac{2}{3}\delta_{ij}\nabla \cdot \bm{u}  \right),
\end{equation}
where we adopt the $\alpha$-viscosity prescription, $\nu = \alpha c_s H$, with $c_s$ and $H$ representing the sound speed and the scale height of the disk, respectively.
$\Phi_{\rm sBH}$ is the gravitational potential of sBHs:
\begin{equation}
    \Phi_{\rm sBH} (\bm{r}) = -\sum_{k=1}^3 \frac{Gm_{k}}{(|\bm{r}-\bm{r}_k|^2+\epsilon^2)^{1/2}},
\end{equation}
where $m_k$, $\bm{r}_k$ and $\bm{r}$ are the sBH mass, location of the $k_{\rm th}$ sBH and the fluid element, and $\epsilon$ is the gravitational softening length. Consistent with \citetalias{Wang2025arXiv}, we adopt a simple gas removal algorithm to model accretion onto the sBH. In this approach, both the mass and momentum are gradually removed from the gas within a radius $r_{\rm sink} = \epsilon$ around the sBH\,\citep[see also][]{LYP2021ApJ906, LYP2021ApJ911}. The removal is implemented phenomenologically such that the surface density satisfies $\dot{\Sigma}/\Sigma = n_{\rm b} \Omega_{\rm b0}$, where $n_{\rm b}$ is a adjustable parameter and $\Omega_{\rm b0} = \sqrt{G(m_1 + m_2)/a_{\rm b0}^3}$ is the initial orbital angular frequency of the BBH. Note that we do not add the mass and momentum of the accreted gas to the sBH. 

We employ the IAS15 integrator\,\citep[][]{Rein2015MNRAS} in the REBOUND framework to compute the dynamical evolution of the three sBHs around SMBH. The gravitational force on the sBHs contributed by the gas can be written as,
\begin{equation} \label{eq:a_gas}
    \bm{a}_{k,\rm gas} = G\sum_{i=1}^{n_x} \sum_{j=1}^{n_y} {\rm d}m_{ij}\, \frac{\bm{r}_{ij}-\bm{r}_{k}}{ (|\bm{r}_{ij}-\bm{r}_{k}|^2 + \epsilon^2)^{(3/2)} },
\end{equation}
where $n_x$ and $n_y$ are the number of fluid cells in the $x$-direction and $y$-direction, ${\rm d}\,m_{ij}$ and $\bm{r}_{ij}$ are the mass and location of the $(i^{\rm th},\, j^{\rm th})$ fluid cell. 
During the BSI process, the pericenter distance between two sBHs may become sufficiently small that GW emission dominates the orbital shrinkage, leading to the formation of a bound binary that subsequently merges rapidly. Therefore, we employ the PN approximation in REBOUND code to model the sBHs' dynamics, to which the relativistic corrections are incorporated as an expansion in powers of \( v/c \)\,\citep[][]{Blanchet2006LRR}:  
\begin{equation} \label{eq:PN}
    \bm{a} = \bm{a}_0 + c^{-2} \bm{a}_2 + c^{-4} \bm{a}_4 + c^{-5} \bm{a}_5 + \mathcal{O}(c^{-6}),
\end{equation}  
where \( \bm{a}_0 \) represents the Newtonian acceleration (0PN order). The terms \( c^{-2} \bm{a}_2 + c^{-4} \bm{a}_4 \) (1PN and 2PN orders) are conservative, contributing to orbital precession without energy dissipation. In addition, the dissipative term \( c^{-5} \bm{a}_5 \) describes the leading-order energy and momentum loss due to GW radiation in this framework. The specific forms of these coefficients can be found in \cite{Blanchet2006LRR}. 

\begin{figure*}
\centering
\includegraphics[scale=0.6]{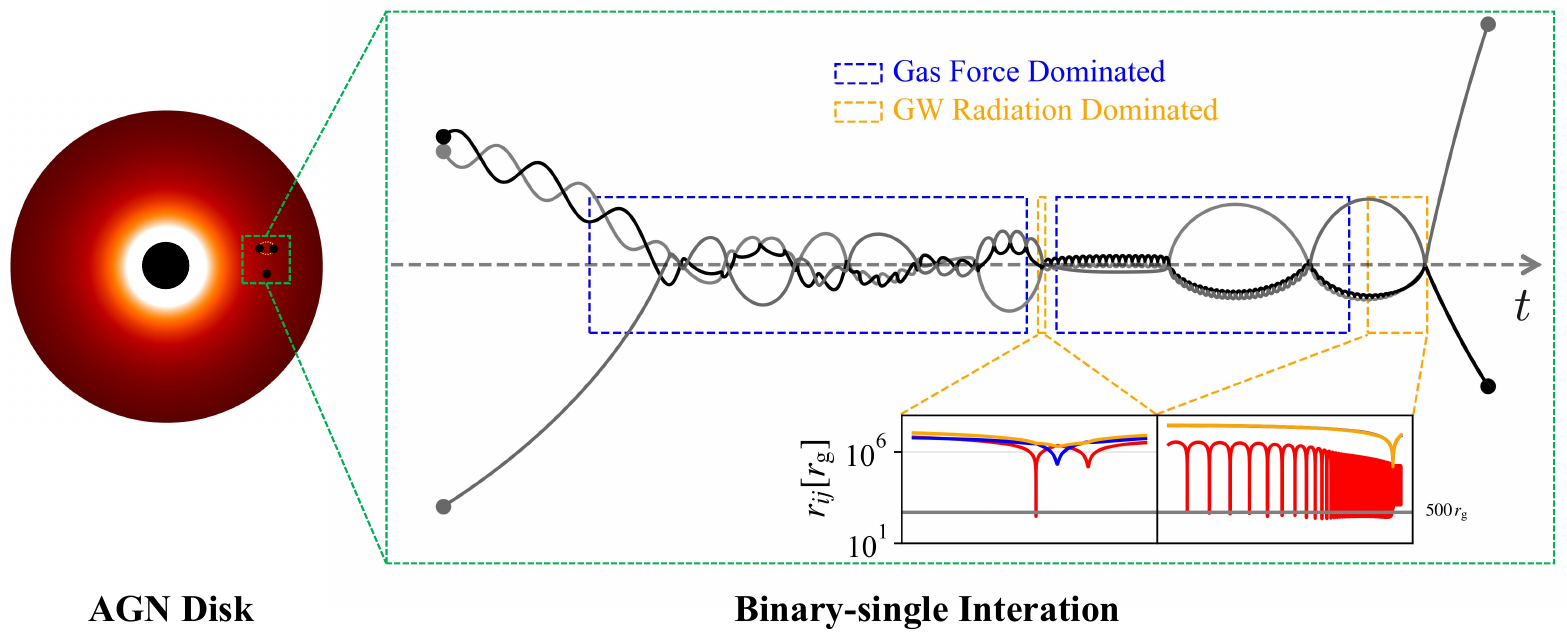}
\caption{Schematic of trajectory during binary–single interaction\,(BSI) in an AGN disk. The energy dissipation of the three-body system is influenced by two dominant mechanisms: (1) gravitational interactions with the surrounding gas, which prevail on large spatial scale and long timescale; and (2) GW radiation, which becomes important at small separations and short timescales—typically when the separation between two sBHs satisfies $r_{ij} \lesssim 500\,r_{\rm g}$. Regions dominated by gas effects are outlined with blue dashed boxes, while red dashed boxes highlight regimes where GW radiation governs the dynamics. The influence of the gas on the three-body system at large scales is coupled with the impact of GW radiation at small scales, which cannot be treated separately.}
\label{fig:cartoon}
\end{figure*}

\begin{table*} 
    \centering
    \fontsize{7}{12}\selectfont    
    \begin{threeparttable}
    \begin{tabular*}{\textwidth}{@{\extracolsep{\fill}}ccccccccccccccc}
        \toprule
        \multirow{2}{*}{Run} & \multirow{2}{*}{$(m_1,m_2,m_3)$} & \multirow{2}{*}{$R_0$} & \multirow{2}{*}{$\Sigma_0$} & \multirow{2}{*}{$L_X\times L_Y$} & \multirow{2}{*}{root grid} & \multirow{2}{*}{$N_{\rm AMR}$}& \multirow{2}{*}{$\epsilon$} & \multirow{2}{*}{$\delta_{\rm min}$} & \multirow{2}{*}{$b$} & \multirow{2}{*}{$a_{\rm b0}$} & \multirow{2}{*}{$\phi_0$} & \multicolumn{3}{c}{$P_{\rm m}$} \\
        \cmidrule(lr){13-15}
          & [$M_\odot$] & $[R_g]$ &[$\frac{M_\bullet R_0^{-2}}{10^4}$] &[$R_{\rm H}^2$] & & &$[R_{\rm H}]$ &$[R_{\rm H}]$ & $[R_{\rm H}]$ & $[R_{\rm H}]$ & & Gas-free & HD & GDF \\
         (1) &(2) & (3) &(4) &(5) &(6) &(7) &(8) &(9) &(10) &(11) & (12)& (13) & (14) & (15)  \\
         \cmidrule(lr){1-15}
         \multirow{9}{*}{I} & \multirow{9}{*}{(50,\,50,\,50)} & \multirow{9}{*}{1000} & \multirow{9}{*}{2.0} & \multirow{9}{*}{$10\times 20$} & \multirow{9}{*}{$512\times 1024$} & \multirow{9}{*}{5} & \multirow{9}{*}{0.012} & \multirow{9}{*}{0.0012} &1.90 & \multirow{9}{*}{0.2} & \multirow{9}{*}{[0,\,$\pi$]} &5/180 &36/180 & 4/180\\
         & & & & & & & & &1.95 & & &0/180 &25/180 & 12/180\\
         & & & & & & & & &2.00 & & &5/180 &14/180 & 5/180\\
         & & & & & & & & &2.05 & & &11/180 &57/180 & 29/180\\
         & & & & & & & & &2.10 & & &17/180 &34/180 & 26/180\\
         & & & & & & & & &2.15 & & &9/180 &40/180 & 22/180\\
         & & & & & & & & &2.20 & & &4/180 &60/180 & 10/180\\
         & & & & & & & & &2.25 & & &5/180 &38/180 & 10/180\\
         & & & & & & & & &2.30 & & &9/180 &27/180 & 23/180\\
        \cmidrule(lr){1-15}
         \multirow{5}{*}{II} & \multirow{5}{*}{(20,\,20,\,20)} & 350 & 0.1 & \multirow{5}{*}{$10\times 25$} & \multirow{5}{*}{$256\times 640$} & \multirow{5}{*}{7} & \multirow{5}{*}{0.006} & \multirow{5}{*}{0.0006} &\multirow{5}{*}{2.10} & \multirow{5}{*}{0.1} & \multirow{5}{*}{[0,\,$\pi$]} &18/180 &3/36 & 18/180\\
         & &1000 &1.0 & & & & & & & & &10/180 &2/36 & 7/180\\
         & &3500 &9.0 & & & & & & & & &11/180 &8/36 & 13/180\\
         & &10000 &15.0 & & & & & & & & &8/180 &11/36 & 16/180\\
         & &35000 &30.0 & & & & & & & & &7/180 &9/36 & 9/180\\
        \bottomrule
    \end{tabular*}\vspace{0cm}
    \end{threeparttable}
    \caption{Simulation setups and results for $M_\bullet=10^8\,M_\odot$ and $H/R=0.01$. Columns:(1) run names; (2) mass of sBHs; (3) location of the shearing-box center; (4) surface density of the ambient AGN disk; (5) size of the simulation domain; (6) root grid; (7) AMR refinement level\,(for example, $N_{\rm AMR} = 5$ means the root grid + 4 finer levels); (8) gravitational softening parameter; (9) minimal size of the cell; (10) impact parameter between the BBH and the single sBH; (11) semi-major axis of the initial BBH; (12) initial orbital phase angle of the BBH with respect to the x-axis, where we take a uniform distribution; (13), (14) and (15) BBH merger probability during BSI process for Gas-free, HD and GDF cases, respectively.   }
    \label{table:1}
\end{table*}

In our coupled HD\,(Athena++) and N-body\,(REBOUND) simulations, both codes use adaptive timestep. To synchronize their evolution, we perform REBOUND integration within each Athena++ timestep until their times match. When sBHs are farther apart than the softening length $\epsilon$, REBOUND's timestep usually exceeds Athena++'s, requiring just one N-body step per hydro step. However, during close encounters, REBOUND's timestep becomes significantly smaller, necessitating multiple N-body steps per hydro step.
Note that, during each close encounter of sBHs, (1) their relative velocities are much higher than the sound speed of the gas, so the gas distribution does not change significantly over the timescale of the encounter;  (2) the gravitational interactions between the sBHs overwhelmingly dominate over the gravitational influence of the surrounding gas. Essentially, the \emph{N}-body dynamical timescale is much shorter than  the gas dynamics timescale when the distance between the two sBHs becomes smaller than $\epsilon$ or even $\delta_{\rm min}$, which is the minimal hydrodynamical grid size $\delta_{\rm min} \gtrsim 10^4\,r_{\rm g}$, where $r_{\rm g} \equiv Gm/c^2$ is the gravitational radius of an sBH. This feature allows us to perform multiple $N$-body timestep within a single Athena++ timestep as an approximation when we deal with the close encounter processes.

\begin{figure}
\centering
\includegraphics[scale=0.49]{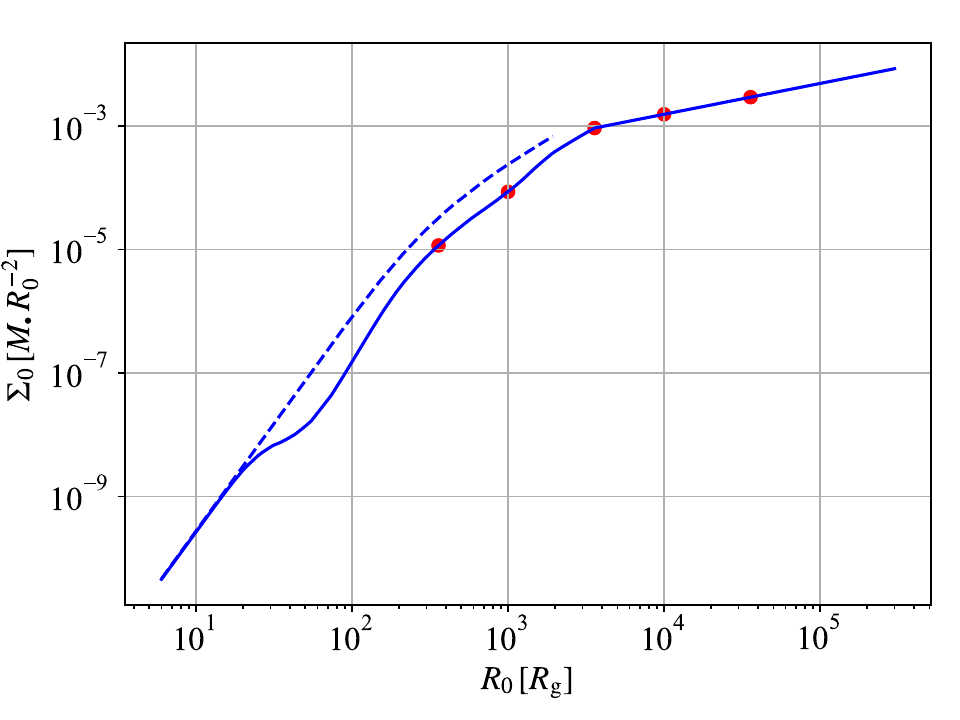}
\caption{The surface density of the AGN disk in the unit of $M_\bullet R_0^{-2}$ for $M_\bullet=10^8\,M_\odot$, $\alpha = 0.1$ and $\dot{M}_\bullet=0.1\,\dot{M}_{\rm Edd}$. The dashed and solid lines represent the SG disk model with $\kappa_1 = 0.34\, {\rm cm}^2\cdot g^{-1} +6.4\times 10^{22}\, {\rm cm}^2\cdot g^{-1}\,\,\rho/(g\cdot {\rm cm}^{-3}) \,(T_{\rm c}/K)^{-3.5}$ and opacity $\kappa_2$ from \cite{Iglesias1996ApJ,Alexander1994ApJ}, respectively. The five dots show the positions of $R_0 = 350\,R_{\rm g}$, $1000\,R_{\rm g}$, $3500\,R_{\rm g}$, $10000\,R_{\rm g}$ and $35000\,R_{\rm g}$ for \emph{Run II}, where $R_{\rm g} \equiv GM_\bullet/c^2$ is the gravitational radius of the SMBH. Note that the boundary between the inner standard thin disk and the outer star-forming disk, $R_{\rm sg}$, is $\sim 3500\,R_{\rm g}$. }
\label{fig:AGN_disk}
\end{figure}

\begin{figure*}
\centering
\includegraphics[scale=0.6]{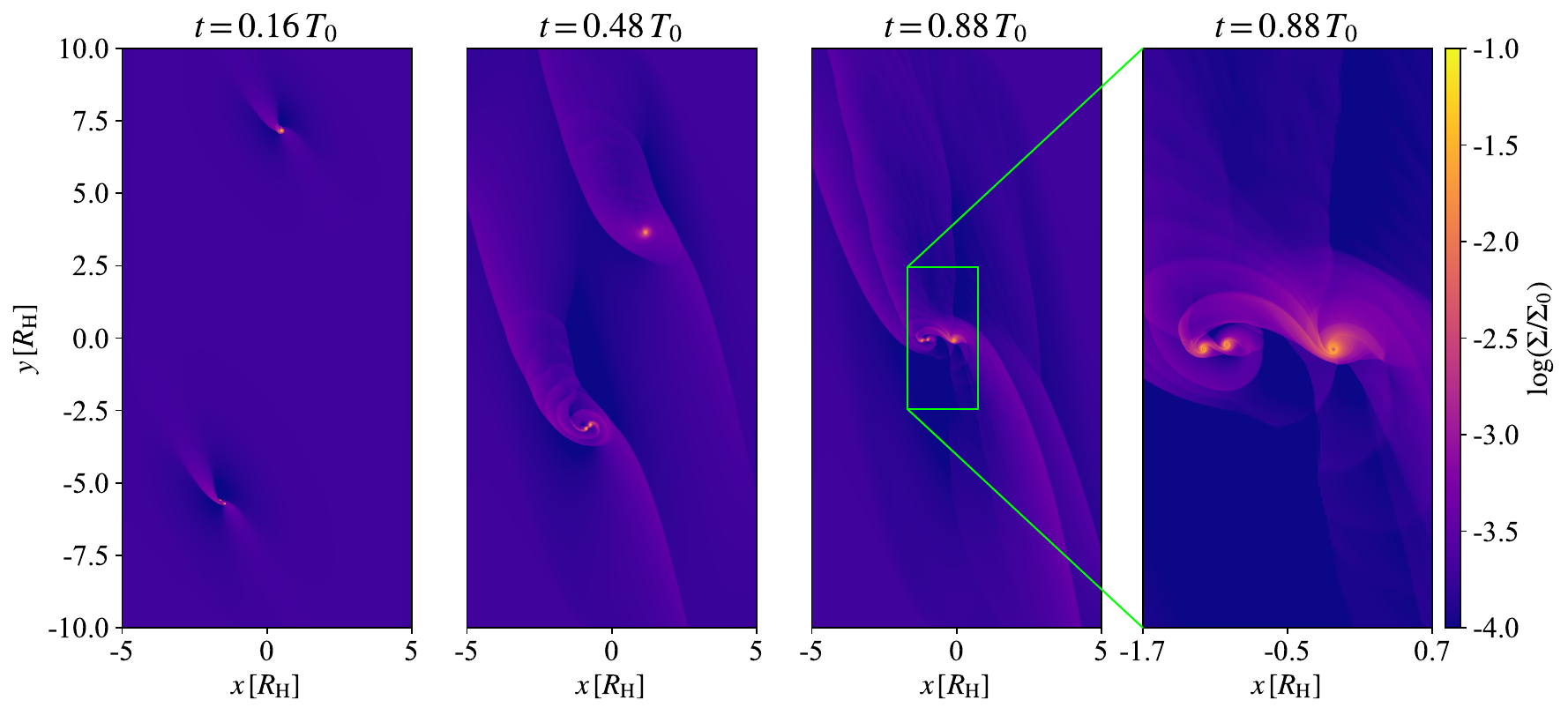}
\caption{Snapshots of the gas surface density for the case with $b=2.1\,R_{\rm H}$ and $\phi_0 = \pi/18$ in Run I, where $T_0=2\pi/\Omega_0$ is the orbital period of the shearing box orbiting around the SMBH The first two panels illustrate the sequential formation of CSDs around the sBHs and the development of their spiral-arm structures. The third panel captures the onset of the BSI, and the fourth panel provides a zoom-in view of the BSI region from the third panel.   }
\label{fig:density_map}
\end{figure*}

\subsection{Energy dissipation of the triple system}

In our model, the orbital dynamics of the three-body system are governed by gravitational forces from the central SMBH, the surrounding dense gas and GW radiation. The role of the SMBH has been discussed in detail in \citetalias{Wang2025arXiv}, which is most significant during the initial close encounter between the BBH and the single sBH. During the subsequent chaotic three-body interactions, however, the effects of gas drag and GW radiation become more dominant. Thus, in this work, we focus on the impact of gas and GW radiation on the dynamical evolution of the triple system. The energy dissipation due to gas drag and GW radiation can be expressed respectively as:
\begin{equation} \label{eq:energy_disssipation}
\begin{split}
    & \varepsilon_{\rm gas} = \sum_{1\leq k<l\leq 3}^3 \frac{m_km_l}{M_{\rm sBH}} (\bm{v}_k-\bm{v}_l)\cdot (\bm{a}_{k,\rm gas}-\bm{a}_{l,\rm gas}),\\
    & \varepsilon_{\rm GW} = \sum_{1\leq k<l\leq 3}^3 \frac{m_km_l}{M_{\rm sBH}} (\bm{v}_k-\bm{v}_l)\cdot (\bm{a}_{k,\rm GW}-\bm{a}_{l,\rm GW}),
\end{split}
\end{equation}
where $m_{k/l}$, $\bm{v}_{k/l}$ are the mass and velocity of sBHs, $\bm{a}_{k/l,\rm gas}$ is given by equation\,\eqref{eq:a_gas} and $\bm{a}_{k/l,\rm GW} = c^{-5} \bm{a}_5$ is the 2.5PN term in equation\,\eqref{eq:PN}.

As illustrated in Figure\,\ref{fig:cartoon}, we present a schematic of the dynamical evolution during BSI in an AGN disk. On large spatial scales and long timescales (blue boxes), the ambient gas influences the orbital trajectories of the sBHs and dominates the energy dissipation processes. When the separation between two sBHs decreases below $\sim 500 r_{\rm g}$ (orange boxes), GW radiation becomes significant on much shorter timescales and influences their orbital dynamics. In such a chaotic interaction process, it will strongly modulate the subsequent dynamical evolution of the triple system.
Therefore, the large-scale gas effects and small-scale GW radiation become intrinsically coupled, indicating that the decoupled treatment may require careful re-examination.

\subsection{The Initial Conditions}
We divide our simulations into two main categories, referred to as \emph{Run I} and \emph{Run II}:

\emph{Run I}: To directly obtain the merger probability during BSI, we explore the parameter space by conducting simulations for 9 distinct impact parameters (\(b = [1.9,\, 1.95,\, 2.0,\, 2.05,\, 2.1,\, 2.15,\, 2.2,\, 2.25,\, 2.3]\,R_{\rm H}\)) and 180 initial phase angles uniformly distributed over \(\phi_0 \in [0, \pi]\), yielding a total of \(9 \times 180\) HD simulation configurations. Consistent with \citetalias{Wang2025arXiv}, the surface density of the ambient disk, $\Sigma_0$, is given by the standard thin disk model\,\citep[][]{Shakura1973} with opacity $\kappa_1 = 0.34\, {\rm cm}^2\cdot g^{-1} +6.4\times 10^{22}\, {\rm cm}^2\cdot g^{-1}\,\,\rho/(g\cdot {\rm cm}^{-3}) \,(T_{\rm c}/K)^{-3.5}$, which is shown by dashed line in Figure\,\ref{fig:AGN_disk}.

\emph{Run II}: To investigate the radial dependence of the BSI in AGN disks, we conducted a series of high-resolution simulations by placing shearing boxes at five distinct radial distances from the central SMBH, which are shown by red dots in Figure\,\ref{fig:AGN_disk}. For each radial position $R_0$, we initialized 36 different orbital phases of the BBH, uniformly distributed across $\phi_0 \in [0,,\pi]$, yielding a total of $5 \times 36$ simulation runs. Since we aim to study the BSI process across both the inner standard thin disk and the outer star-forming disk regions, we adopt  the SG model given by \cite{Sirko2003}, while the opacity parameter $\kappa_2$ is derived from the opacity tables provided in \cite{Iglesias1996ApJ} and \cite{Alexander1994ApJ}. In this case, the boundary between the inner
standard thin disk and the outer star-forming disk, $R_{\rm sg}$, is about $3500\,R_{\rm g}$.

In this work, we assume an isothermal equation of state $P = \Sigma c_s^2$ with sound 
speed $c_s = H\Omega_0$. All simulations adopt $M_\bullet=10^8\,M_\odot$, disk aspect ratio $H/R=0.01$, viscosity coefficient $\alpha = 0.1$ and gas removal rate $n_b = 5.0$. Notably, our simulations maintain a fixed y-direction\,(azimuthal) separation of $16R_{\rm H}$ independent of x-direction\,(radial) separation $b$.
To ensure the gas reaches a steady state before BSI and to highlight its impact on the BSI process, we disable gas gravity on the sBHs until the separation between the BBH's center of mass and the single sBH falls below $R_{\rm H}$. We summarize the other key parameters and results presented in this paper in Table\,\ref{table:1}. In Figure\,\ref{fig:density_map}, we present three snapshots of the gas morphology before BSI to more clearly illustrate our initial setup, where $T_0=2\pi/\Omega_0$ is the orbital period of the shearing box orbiting around the SMBH. Note that the gas morphology during BSI has already been shown in \citetalias{Wang2025arXiv} and is therefore not repeated here.


\section{Results} \label{sec:results}

Following \citetalias{Wang2025arXiv}, we categorize the end states of the BSI into the following four types:\\
\indent $\bullet$ \textbf{BS}: binary + single sBH state, in which three sBHs undergo chaotic triple interactions and eventually reconfigure into a BBH and an isolated sBH. We denote the state in which sBH${1(/2/3)}$ and sBH${2(/3/1)}$ form a binary as BS12(/23/13). \\
\indent $\bullet$ \textbf{ST}: stable triple state, where a single sBH orbits at a relatively large distance around a compact BBH, and this system remains stable until the end of our simulation.\\
\indent $\bullet$ \textbf{Ion}: ionization state, where the three-body system disperses into three single sBHs.\\
\indent $\bullet$ \textbf{GW}\,(or BBH merger during BSI): Two sBHs shrink to a final separation of less than $10\,r_{\rm g}$ due to GW radiation during the BSI process, which means that this merger event will happen before the third sBH falls back.
\\

Since we are particularly interested in the BBH merger events during the BSI process in this work, we will highlight the GW end state in the following discussions. The BBH merger probability during BSI can be approximately estimated as  
\begin{equation}  \label{eq:3b_merger_rate}
P_{\rm m} \approx N_{\rm enc} \times P(r_{\rm p} < r_{\rm c}),
\end{equation}  
where \( N_{\rm enc} \) is the number of binary-single encounters during the BSI process. The $N_{\rm enc}$ is modulated by the tidal force exerted by the central SMBH as $N_{\rm enc} = 20\times (1-2a_{\rm b}/R_{\rm H})^{2}$\,\citep[see][]{Fabj2024MNRAS}. The \( P(r_{\rm p} < r_{\rm c}) \) represents the probability that the BBH's pericenter distance falls below a critical pericenter \( r_p=r_{\rm c} \). This probability was estimated in the Gas-Free environment in~\cite{Fabj2024MNRAS} through equating two time-scales: the BBH merger timescale in the high eccentricity limit  \( T_{\rm m} \)\,\citep[][]{Peters1964}, and  the typical Keplerian period of the single object orbiting around the BBH \( T_{\rm bs} \):
\begin{equation}\label{eq:timescale}
\begin{split}
    T_{\rm m} &\approx \frac{768}{425}\frac{2^{7/2}5\,c^5\,a_{\rm b}^{1/2}\,r_{\rm p}^{7/2}}{256\,G^3\,m_1 m_2 (m_1 + m_2)}, \\
    T_{\rm bs} &\approx 2\pi \sqrt{\frac{a_{\rm bs}^3}{G M_{\rm sBH}}},
\end{split}
\end{equation}
where \( a_{\rm b} \) is the semi-major axis of the BBH, \( a_{\rm bs} \) is the characteristic binary-single separation. Assuming \( a_{\rm bs} \sim a_{\rm b} \) and equal-mass sBHs (\( m_1 = m_2 = m_3 \)), we obtain an estimate for the critical pericenter as:
\begin{equation}\label{eq:rc}
r_{\rm c} \approx 2.3\,r_{\rm g} \times \left(\frac{a_{\rm b}}{r_{\rm g}}\right)^{2/7},
\end{equation}
Since the eccentricity of BBHs in the coplanar BSI process follows a superthermal distribution \( P(e) = e/\sqrt{1-e^2} \), we can further express the term \( P(r_{\rm p} < r_{\rm c}) \) as:
\begin{equation} \label{eq:ec} 
    P(r_{\rm p} < r_{\rm c}) = P(e>e_{\rm c}) = \sqrt{1-e_{\rm c}^2}
\end{equation}
with $e_c$ as a function of the BBH semi-major axis $e_{\rm c}(a_b) = 1-2.3(r_{\rm g}/a_{\rm b})^{5/7} $, where we have used the relation of $r_{\rm p} = a_b(1-e)$.

To account for the gas influence on the probability $P(r_p<r_c)=P(e>e_c)$, \cite{Rowan2025arXiv} proposed the same form of the probability expect that the semi-major axis $a_b$ in $e_c(a_b)$ is shorten by the gas effect. \cite{Rowan2025arXiv} defined an enhancement factor of $P(r_p<r_c)$ as $\eta= P(e>e_{\rm c}(a_{\rm f}))/P(e>e_{\rm c}(a_{\rm b0}))$, where $a_{\rm b0}$ and $a_{\rm f}$ represent the spatial scales of the initial and final triple systems, respectively. Through this approach, they found that $\eta\approx 3.5$ and $8$ if gas can shrink the final triple system to the size of gravitational softening length $\epsilon$ and the minimal resolution cell $\delta_{\rm min}$, respectively. 

While this analytical approach is intriguing, additional factors need to be taken into account for a better merger-rate estimation. For example, (1) the encounter times $N_{\rm enc}$ can also significantly affect the probability $P_m$; (2) during the three-body interaction process, the size of the three-body system is shrinking, hence using the end-state scale throughout the entire process 
likely overestimates the enhancement. To circumvent these issues, in the following, we present the BBH merger probability directly obtained from our coupled HD and Post-Newtonian \emph{N}-body simulations.

\subsection{Run I: BBH Merger Probability during BSI}
\begin{figure}
\centering
\includegraphics[scale=0.46]{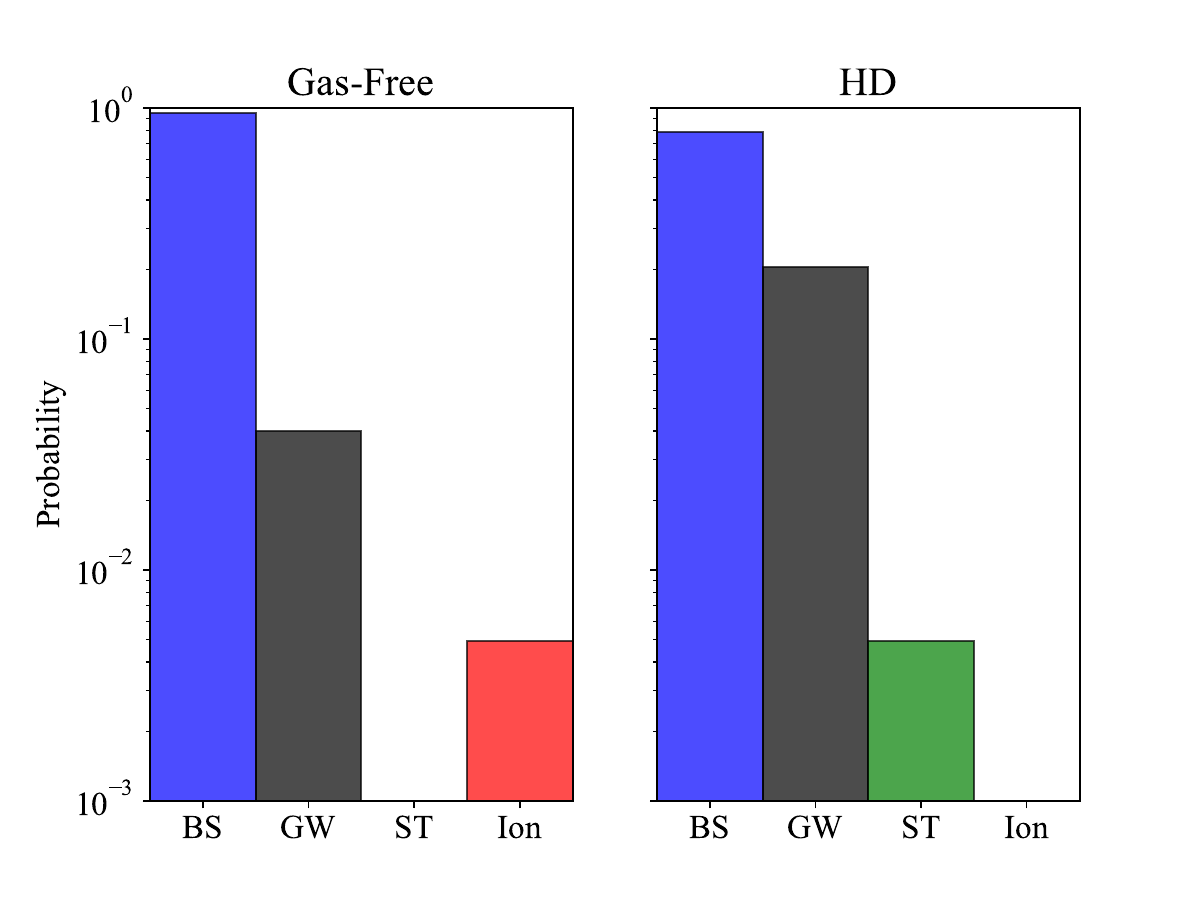}
\caption{The statistical probabilities of the end states for $9\times 180$ HD simulations in \emph{Run I}. The left and right panels show the Gas-Free and HD cases, respectively. The presence of gas increases the GW state from 4.0\% to 20.4\%. }
\label{fig:hist}
\end{figure}

In \emph{Run I}, we conduct \(9 \times 180\) HD simulations to examine how gas affects the BBH merger probability during BSI. In Figure\,\ref{fig:hist}, we present the statistical distribution of the various end states.
The majority of end states fall into the BS state. In the Gas-Free case, a small fraction of triple systems are disrupted into three single sBHs (i.e., the Ion state) after the BSI process, and no ST state is observed.
Conversely, in the HD case, a small fraction occupies the ST state, while no Ion state appears. Importantly, compared to the Gas-Free scenario, the presence of gas significantly increases the occurrence of GW state. We find that dense gas in the AGN disk enhances the fraction of the GW state by $\sim 5$ times\,(from 4\% to 20.4\%).

\begin{figure*}
\centering
\includegraphics[scale=0.6]{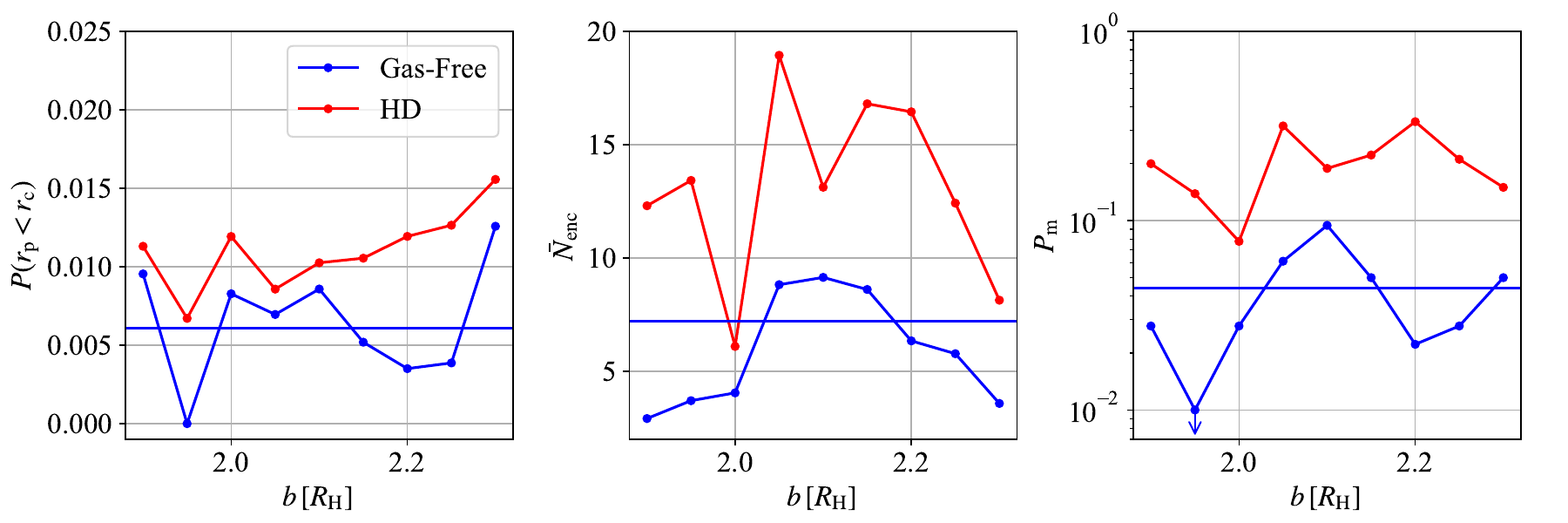}
\caption{ 
\textbf{The left panel}: the probability, \(P(r_{\rm p}<r_{\rm c})\), that the pericenter of these temporary BBHs during the BSI processes falls below a critical value \(r_{\rm c}\) under different impact parameters $b$. The blue and red dots represent the cases of the Gas-Free and HD, respectively. 
\textbf{The middle panel}: the average number of encounters between the single sBH and the BBH during the BSI process\,(or the average occurrence number of the temporary BS states). \textbf{The right panel}: the BBH merger probability during BSI for different impact parameters $b$, which are obtained directly from our simulations. For the Gas-Free case, the analytical predictions suggested that
\(P(r_{\rm p}<r_{\rm c})\approx 0.0061\),  $\bar{N}_{\rm enc}\approx 7.2$ and $P_{\rm m} = 4.4\%$ for $a_{\rm b} = R_{\rm H}/5$, which are shown by blue horizontal lines in all the three panels.} 
\label{fig:encounter_number}
\end{figure*}

During the BSI process, the triple system undergoes multiple temporary BBH formations, each accompanied by a single sBH — collectively referred to as temporary BS states. As indicated by Equation\,\eqref{eq:3b_merger_rate}, the BBH merger probability during BSI, \( P_{\rm m} \), depends on two key factors: (1) the probability of these temporary BBHs having a pericenter distance smaller than the critical radius \( r_{\rm c} \), \( P(r_{\rm p} < r_{\rm c}) \), and (2) the total number of temporary BS states \( N_{\rm enc} \).  
To investigate the mechanisms responsible for the enhanced probability, we obtain the probability \( P(r_{\rm p} < r_{\rm c}) \) and the average number of binary-single encounters $\bar{N}_{\rm enc}$ directly from our simulations for each impact parameter \( b \). We present the results for both the Gas-Free case\,(blue dots) and the HD case\,(red dots) in the left and middle panels of Figure\,\ref{fig:encounter_number}. In the right panel of Figure\,\ref{fig:encounter_number}, we present the BBH merger probability from our simulations. For the Gas-Free scenario, analytical theory predicts \( P(r_{\rm p} < r_{\rm c}) \approx (a_{\rm b0}/2r_{\rm g})^{-5/14} \approx 0.61\% \), and \( N_{\rm enc} \sim 20 \times (1 - 2a_{\rm b0}/R_{\rm H})^2 \approx 7.2 \) for \( a_{\rm b0} = R_{\rm H}/5 \)\,\citep[see][]{Fabj2024MNRAS}. Consequently, the analytically expected BBH merger probability in the coplanar and Gas-Free case is approximately \( P_{\rm m} \sim 4.4\% \). 
The analytical prediction is shown as blue horizontal lines in Figure\,\ref{fig:encounter_number}, which is consistent with our \emph{N}-body simulations for the Gas-Free case.

As shown in Figure\,\ref{fig:encounter_number}, both $P(r_{\rm p} < r_{\rm c})$ and $\bar{N}_{\rm enc}$, are enhanced by a factor of approximately 2, ultimately leading to a $\sim$5-fold increase in the BBH merger probability. This indicates that both the shrinkage of the triple system and the increased number of binary–single encounters contribute significantly to the merger probability. Focusing solely on the enhancement of $P(r_{\rm p} < r_{\rm c})$, as in \citet{Rowan2025arXiv}, would underestimate the gas effect. Notably, the gas-induced increase in $\bar{N}_{\rm enc}$ slightly exceeds that in $P(r_{\rm p} < r_{\rm c})$, highlighting the important role of repeated binary-single encounters in boosting merger probability. 

Compared to our results, the simulations of \cite{Rowan2025arXiv} show a stronger gas-induced shrinkage of the triple system's scale. In their simulations, 24 of 65 BSI end states are classified as hardened chaotic encounters, in which case the size of the three-body system continues to shrink until it reaches the gravitational softening length $\epsilon$ or the minimum cell size $\delta_{\rm min}$. Consequently, they find that gas can enhance $P(r_{\rm p} < r_{\rm c})$ by a factor of 3.5 and 8, significantly larger than the factor of 2 increase observed in our simulations. Two main factors contributing to this difference are: (1) they adopt a non-accreting sBH model; and (2) their simulations focus on BSI processes occurring at larger distances from the central SMBH. Gas accretion onto sBHs in AGN disks can significantly alter the structure of their CSDs, which in turn affects the energy dissipation during the BSI process. We will further discuss this accretion issue in Section\,\ref{sec:diss_lim}. In the following subsection, we will investigate how the impact of gas on the BSI process varies with distance from the central SMBH within the AGN disk.




\begin{figure*}
\centering
\includegraphics[scale=0.61]{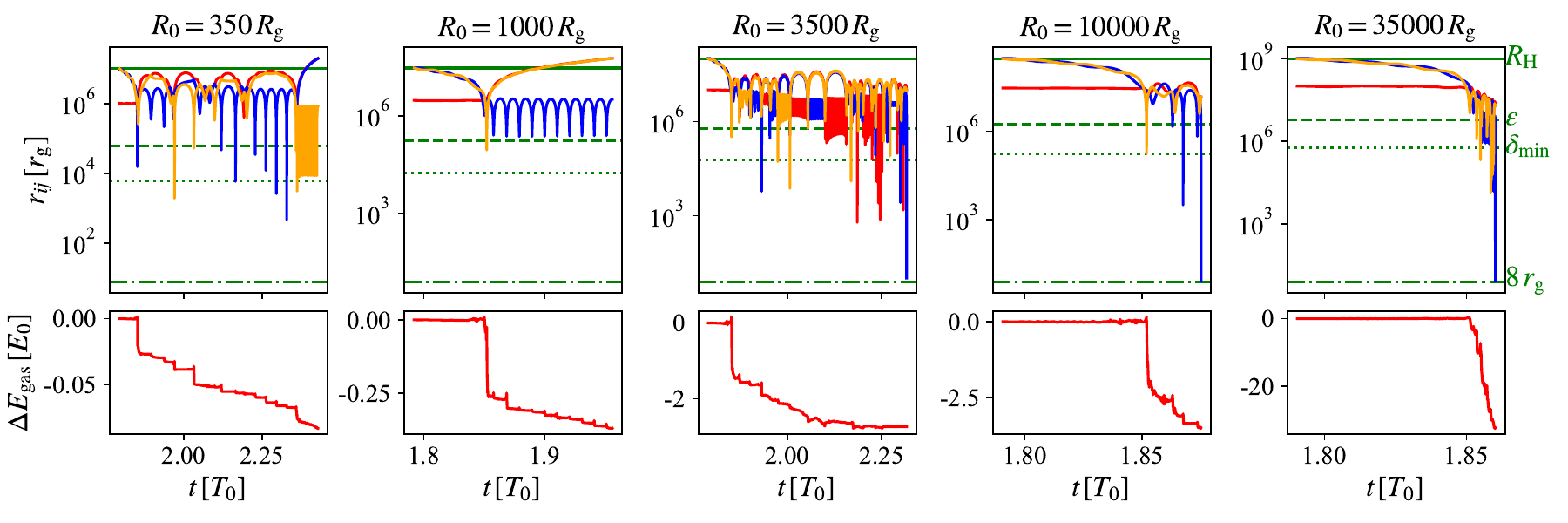}
\caption{Evolution of sBH separations and energy dissipation in three-body systems during the BSI process for \emph{Run II} with the BBH's initial orbital phase \( \phi_0 = 0 \). In the top panels, the red, blue, and orange lines indicate the separations \( r_{12} \), \( r_{23} \), and \( r_{13} \), respectively. The four horizontal green lines (from top to bottom) mark the Hill radius \( R_{\rm H} \), the softening length \( \epsilon \), the minimum grid cell size \( \delta_{\rm min} \), and the merger radius \( 10\,r_{\rm g} \). The bottom panels show the cumulative gas-induced energy dissipation, defined as \( \Delta E_{\rm gas} \equiv \int_0^t \varepsilon_{\rm gas} \,{\rm d}t \). Energy is normalized by \( E_0 \), the initial total energy of the three-body system when the single sBH is at a distance \( R_{\rm H} \) from the BBH’s center of mass. The characteristic timescale \( T_0 = 2\pi/\Omega_0 \) corresponds to the orbital period in the shearing-box frame.}
\label{fig:RunI_distance}
\end{figure*}

\begin{figure*}
\centering
\includegraphics[scale=0.49]{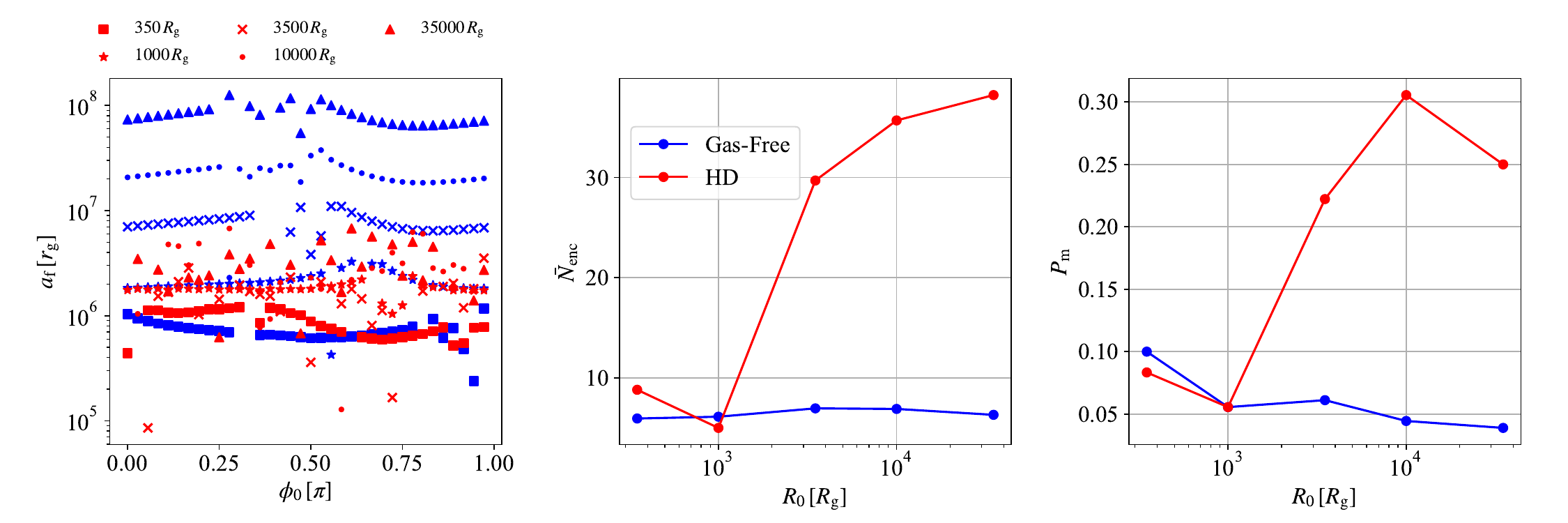}
\caption{Results of \emph{Run II} with varying initial radii \( R_0 \) and BBH orbital phases \( \phi_0 \). 
(a) The semi-major axis $a_{\rm f}$ of the end-state BBHs. Red symbols—squares, stars, crosses, dots, and triangles—represent HD simulations, while blue symbols indicate the corresponding Gas-Free cases for comparison.   
(b) The average number of binary-single encounters as a function of \( R_0 \),
(c) The probability of BBH merger during BSI as a function of \( R_0 \). }
\label{fig:RunI_nenc}
\end{figure*}

\subsection{Run II: Radial Dependence}


Previous studies have suggested that the outer boundary of AGN disks can extend to parsec scales, where radiation pressure and feedback from supernovae of in situ formed stars help maintain disk stability\,\citep[e.g.,][]{SG2003MNRAS,TQM2005ApJ}. These star-forming regions naturally host a population of sBHs, originating either from the evolution of in situ stars or through captures from the surrounding nuclear star cluster. Using one-dimensional \emph{N}-body simulations, \citet{Tagawa2020ApJ} investigated the formation and evolution of BBHs in such environments. They found that although BBHs tend to form in the outer regions of the disk, efficient orbital migration drives them rapidly inward\,\citep[see Figures 8 and 9 in][]{Tagawa2020ApJ}, implying that BSIs can occur throughout the entire disk. To explore how the BSI process depends on the radial location within the AGN disk, we further carry out \emph{Run II}, which spans a wide radial range from $0.1\,R_{\rm sg}$ to $10\,R_{\rm sg}$.

In Figure\,\ref{fig:RunI_distance}, we show the evolution of sBHs' separation and energy dissipation of the three-body system during the BSI process for \emph{Run II} with initial BBH's orbital phase $\phi_0 = 0$. In the upper panels, the red, blue, and orange lines show the separation between sBHs $r_{12}$, $r_{23}$ and $r_{13}$, respectively. The four horizontal green lines from top to bottom represent the Hill radius $R_{\rm H}$, the gravitational softening parameter $\epsilon$, the minimal cell size $\delta_{\rm min}$ and the merger radius $10\,r_{\rm g}$. The strength of gas's role in the BSI process can be characterized by its energy dissipation in the triple system, which can be given by the time integration of the gas power
\begin{equation}
\Delta E_{\rm gas}\equiv \int_0^t \varepsilon_{\rm gas} \,{\rm d}\,t,
\end{equation}
where $\varepsilon_{\rm gas}$ can be found in Equation\,\eqref{eq:energy_disssipation}. We show the energy dissipation by gas in the unit of $E_0$ in the bottom panels, where $E_0$ is defined as the initial total energy of the three-body system when the separation between the single sBH and the BBH's center of mass equals the Hill radius $R_{\rm H}$. 

We find that as \( R_0 \) increases, the energy dissipation induced by the gas becomes progressively more significant. The reason is as follows: as shown in column 4 of Table\,\ref{table:1} or Figure\,\ref{fig:AGN_disk}, the total gas mass enclosed within the Hill sphere of the three-body system increases with $R_0$, approximately scaling as \( \Sigma_0 R_{\rm H}^2 \propto \Sigma_0 R_0^2 \). This implies that in the outer regions of the AGN disk, the mass of surrounding gas within their Hill sphere becomes more substantial, thereby enhancing its dynamical influence on the triple system. Consequently, the role of gas in shaping the outcome of BSI grows increasingly important at larger disk radii.

In Figure\,\ref{fig:RunI_nenc}, we show all results of our $5\times 36$ coupled HD and \emph{N}-body simulations represented by red symbols. For comparison, we present the results of the Gas-Free case with blue symbols. Figure\,\ref{fig:RunI_nenc}(a) displays the semi-major axis $a_{\rm f}$ of the surviving BBH after the BSI\,(BS state). Different marker shapes represent distinct positions \( R_0 \) on the AGN disk. Figure\,\ref{fig:RunI_nenc}(b) shows the average number of binary-single encounters during BSI across 36 simulations with different initial phases $\phi_0$ at various radii $R_0$, while Figure\,\ref{fig:RunI_nenc}(c) shows the BBH merger probability during BSI as a function of $R_0$.

For the Gas-Free case, the BBH merger probability during BSI gradually decreases with increasing \(R_0\), as anticipated from Equation\,\eqref{eq:3b_merger_rate}. The initial semi-major axis of the BBH is set to \(a_{\rm b0} = 0.1\,R_{\rm H}\), so the number of binary-single encounters, \(N_{\rm enc}\), remains constant with increasing \(R_0\). However, the critical eccentricity \(e_{\rm c}\) increases with \(R_0\), leading to a decrease in \(P(r_{\rm p} < r_{\rm c})\). In other words, at larger distances from the central SMBH, the overall scale of the triple system becomes larger, making BBH mergers more difficult.

For the HD case, the three-body system undergoes significant contraction during the BSI process compared to the Gas-Free case. This contraction effectively compensates for the reduced BBH merger probability that would otherwise result from the larger spatial scale of triple systems at greater $R_0$. As a result, the characteristic scale of the end-state BBHs exhibits only a weak dependence on the radial location $R_0$\,(as shown in Figure \,\ref{fig:RunI_nenc}(a)). Moreover, the number of binary–single encounters during the BSI process increases with increasing $R_0$ due to the stronger gas drag, further enhancing the BBH merger probability in the outer regions of the AGN disk. Overall, the BBH merger probability during BSI can be enhanced by factors of 3.7, 6.9, and 6.4 at $R_0 = 3500$, $10000$, and $35000\,R_{\rm g}$, respectively, reaching values as high as 20\%–30\%, while gas has little influence on the BSI process in the inner AGN disk ($\leq 1000\,R_{\rm g}$). 



\subsection{Compare to Gas Dynamical Friction}
High-resolution HD simulations are computationally expensive, with each simulation requiring 100-400 CPU hours (totaling $\sim$ 200,000 CPU hours for our full suite of 1,800 simulations). This substantial computational cost makes extensive parameter space exploration challenging. As an alternative approach, we therefore also employ GDF modeling\,\citep[see][]{Ostriker1999ApJ} to investigate gas effects. In this subsection, we will compare the statistical results of GDF+$N$-body and HD+$N$-body simulations to verify the feasibility of GDF.

In the GDF model, the acceleration of the gas acting on the sBH can be given by
\begin{align}
  & \textbf{a}_{\rm GDF} = \frac{f(x)}{x^3} \frac{\textbf{v}-\textbf{v}_{\rm K}}{\tau_0}, \\ 
  f(x) &= 
  \begin{cases}
    \frac{1}{2} {\rm ln} \left( \frac{1+x}{1-x}\right) -x, & 0< x \leq 1, \\
    \frac{1}{2}{\rm ln} \left(x^2-1\right) + 3.1, & x >1,
  \end{cases}
\end{align}
where $\textbf{v}$ is the velocity of the sBH, $\textbf{v}_{\rm K}$ is the local Keplerian velocity of the disk gas around the center SMBH, $x \equiv |\textbf{v}-\textbf{v}_{\rm K}| / c_s$ is the Mach number of the relative velocity between the sBH and the background disk gas. The characteristic damping timescale  $\tau_0=c_s^3/(4\pi G^2\rho_0 m)$ is determined by the background gas density $\rho_0\,(=\Sigma_0/2H)$, the sound speed $c_s$ and mass $m$ of the sBH.

We show the BBH merger probability during BSI for the GDF case in the last column of Table\,\ref{table:1}. Our results indicate that the GDF approximation fails to reproduce outcomes consistent with those from HD simulations.
The primary reason is that GDF cannot adequately account for the energy dissipation of the three-body system during close encounters of two sBHs.

\section{Discussions} \label{sec:discussion}

In this section, we discuss several key topics, including the significance of conducting direct PN simulations, the conditions of BSI occurrence, potential observational features of BSI in AGN disk, and the limitations of our simulations.

\subsection{Importance of Directly PN Simulations}\label{sec:importance_of_PN}

The BBH merger probability can also be estimated without incorporating PN corrections by conducting HD+Newtonian \emph{N}-body simulations and tracking the pericenter distances $r_{\rm p}$ during the BSI process. Based on Equation\,\eqref{eq:timescale}, a merger can be assumed to occur when $r_{\rm p} < r_{\rm c}$. However, during chaotic triple interactions, the effects of GW emission and gaseous forces are intrinsically coupled with the dynamical evolution of the system and cannot be accurately captured by a decoupled treatment. When two sBHs approach within a close-encounter distance of $500\,r_{\rm g}$, energy loss due to GW radiation begins to influence their orbital dynamics and the number of binary–single encounters\,\citep[see also][]{Samsing2014ApJ}. However, the critical pericenter distance for a BBH merger, $r_{\rm c}$, is smaller than $200\,r_{\rm g}$. As a result, for BSI processes involving close encounters where $r_{\rm c} < r_{\rm p} < 500\,r_{\rm g}$ (i.e., encounters close enough to be affected by GW emission but not close enough to result in immediate merger), Newtonian and post-Newtonian treatments yield divergent outcomes. 

\begin{figure}
\centering
\includegraphics[scale=0.59]{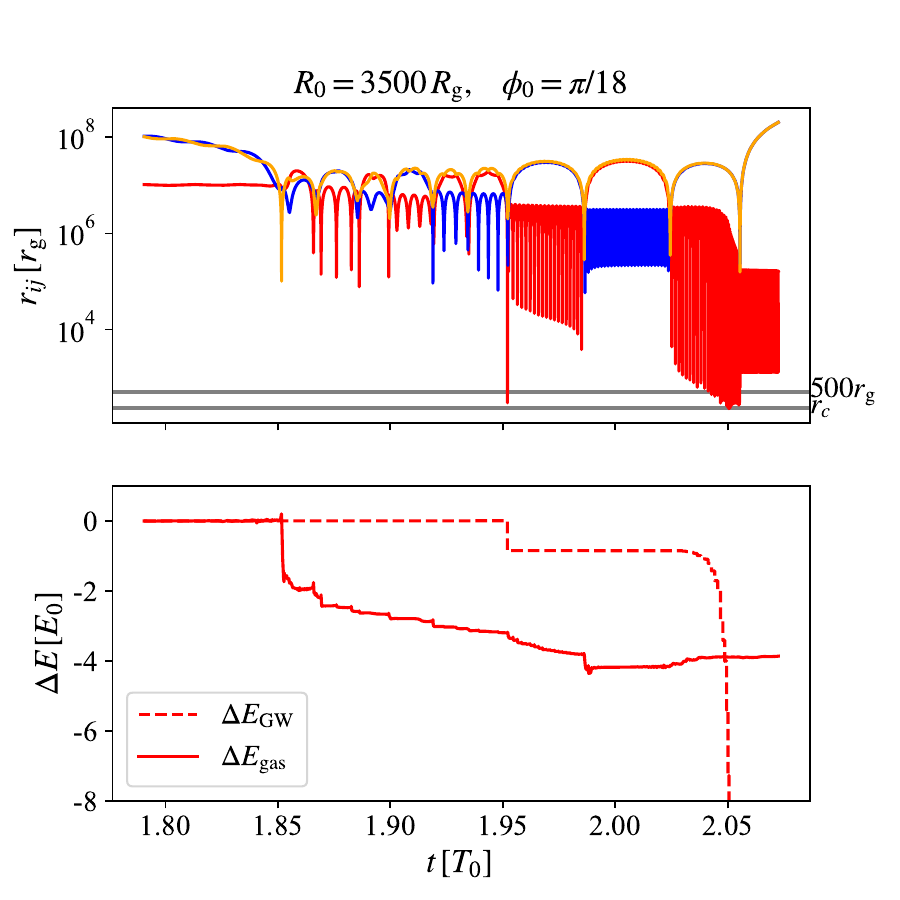}
\caption{Similar to Figure\,\ref{fig:RunI_distance}, but for a BBH with an initial orbital phase of $\phi_0 = \pi/18$. The region between the two gray lines in the top panel denotes the range where $r_{\rm c} < r_{ij} < 500\,r_{\rm g}$. In the bottom panel, the solid and dashed lines represent the energy dissipation of the triple system due to gas and GW radiation, respectively. }
\label{fig:example}
\end{figure}

We present an example of such a case in Figure\,\ref{fig:example}. As shown in the top panel, the separation between $m_1$ and $m_2$ repeatedly falls within the range $r_{\rm c} < r_{12} < 500\,r_{\rm g}$. During these episodes, the energy dissipation due to GW radiation becomes comparable to, or even exceeds, that caused by the gaseous environment, which can significantly alter the subsequent orbital evolution of the triple system. In our simulations, the fraction of such events is approximately 10\%, which is non-negligible given that the overall merger probability $P_{\rm m}$ is only around 20\%.



\subsection{Conditions of BSI Occurrence}

In this work, we calculate the BBH merger probability in AGN disks under the condition that BSI occurs. However, the observable merger rate is determined by $\Gamma \propto n_{\rm AGN} \, n_{\rm BBH} / \tau_{\rm enc}$, where $n_{\rm AGN}$, $n_{\rm BBH}$, and $\tau_{\rm enc}$ denote the number density of AGN, the number density of BBHs in AGN disks, and the timescale of BSI occurrence, respectively.

Previous studies have shown that dense gas can accelerate the formation of BBHs\,\citep[e.g.,][]{LJR2023ApJ, Rowan2023MNRAS, Whitehead2024MNRAS}, suggesting that the BBHs' number density in AGN disks, $n_{\rm BBH}$, could be higher than in other environments\,\citep[][]{Tagawa2020ApJ, Rowan2024arXiv}. However, another critical question is whether a dynamically formed BBH in the AGN disk will merge before experiencing another encounter with a single sBH. The subsequent evolution of these newly formed BBHs in AGN disks remains highly controversial. For example, three-dimensional simulations show that equal-mass BBHs with separations greater than 10\% of their Hill radius contract, while BBHs expand their orbits at smaller separations\,\citep[][]{Dempsey2022ApJ}. Further simulations of eccentric BBHs show that retrograde binaries may undergo runaway eccentricity growth, whereas prograde systems experience slower decay in both semi-major axis and eccentricity\,\citep[][]{Calcino2024ApJ}. \cite{Dittmann2025arXiv} subsequently expanded the parameter space to include eccentricity and unequal-mass BBH systems, similarly observing that retrograde BBHs exhibit faster orbital decay compared to their prograde counterparts. For these prograde BBHs, BSI may offer an alternative channel to drive them toward a merger. Indeed, in the one-dimensional \emph{N}-body simulations of \citet{Tagawa2020ApJ}, BSI was found to play a significant role in shrinking BBH orbits.
 
The timescale of BSI occurrence in AGN disk can be given by $\tau_{\rm enc} = 1/(n_{\rm sBH} \sigma v_{\rm rel}) \sim 1/(3n_{\rm sBH}\Omega_0 R_{\rm H}^2)$, where $n_{\rm sBH}$ is the surface number density of sBHs embedded in AGN disk, $\sigma\sim R_{\rm H}$ is the cross section for the encounter in coplanar case and $v_{\rm vel}\sim \frac{3}{2}\Omega_0 \, 2R_{\rm H}$ is the relative velocity between the BBH and single sBH. In AGN disk, the combination of sBH migration, capture from outside the disk, and in-situ star formation leads to $n_{\rm sBH}$ significantly higher than that in stellar clusters. Additionally, migration traps in the disk can also cause sBHs accumulation, further enhancing the local density\,\citep[e.g.,][]{Masset2003ApJ,Lyra2010ApJ,McKernan2012MNRAS,Bellovary2016ApJ,Secunda2019ApJ}. Consequently, the timescale for BSI occurrences in AGN disks ($\tau_{\rm enc}$), aligning with the BBH dynamical formation timescale, is substantially shorter than the duty cycle of AGN\,\citep[e.g.,][]{Tagawa2020ApJ,McKernan2024arXiv,Rowan2024arXiv}. 



\subsection{Observational Features}

\begin{figure}
\centering
\includegraphics[scale=0.57]{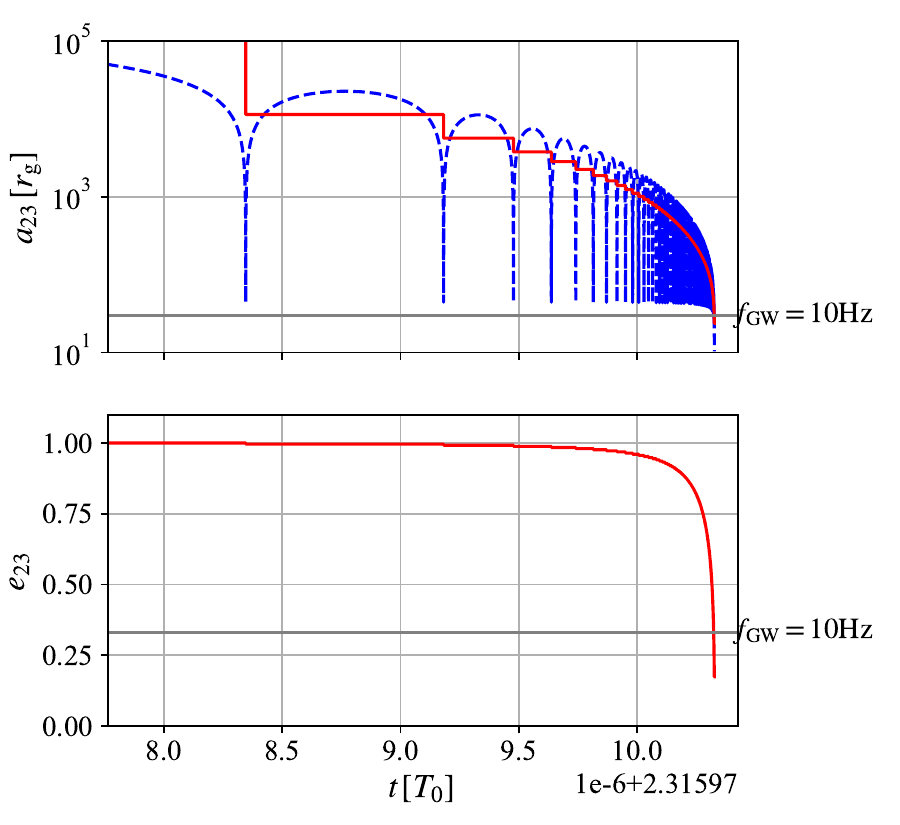}
\caption{ The evolution of semi-major axis and eccentricity during the final stage for the case of $R_0 = 3500\,R_{\rm g}$ and $\phi_0 = 0$\,(as shown in the third panel of Figure\,\ref{fig:RunI_distance}). The blue dashed line in the top panel shows the separation $r_{23}$ between $m_2$ and $m_3$, while the red solid lines show the evolution of their orbital semi-major axis $a_{23}$ (top panel) and eccentricity $e_{23}$ (bottom panel). The two grey lines correspond to the BBH semi-major axis and eccentricity at a GW number of 10 Hz, respectively. Note that both the $a$ and $e$ are estimated by using the geometric method: $a = (r_{\rm a} +r_{\rm p})/2$ and $e = 1-r_{\rm p}/a$.   }
\label{fig:ae_GW}
\end{figure}

$\bullet$ \emph{Eccentric Mergers}: During the BSI process, BBH mergers naturally exhibit high eccentricities. In Figure\,\ref{fig:ae_GW}, we zoom in on the final stage of the third panel in Figure\,\ref{fig:RunI_distance} as an example of a BBH merger during BSI. We show the evolution of their orbital semi-major axis \( a_{23} \) (top panel) and eccentricity \( e_{23} \) (bottom panel) as red solid lines. The blue line in the top panel represents the instantaneous separation \( r_{23} \) between \( m_2 \) and \( m_3 \). The gray horizontal lines indicate the values of the semi-major axis and eccentricity when the frequency of GW reaches 10\,Hz. Remarkably, the BBH retains a relatively high eccentricity (\( e \sim 0.3 \)) even at the 10\,Hz GW frequency.

$\bullet$ \emph{Double GW Mergers}: Following the merger of two sBHs in the BSI process, the resulting BBH remnant may experience a subsequent merger with the third sBH, which is called  double GW mergers\,\citep[e.g.,][]{Samsing2018MNRAS,Samsing2019MNRASb}. If the time interval between these two GW events is sufficiently short, LIGO-like detectors may be able to resolve both signals. Many studies have investigated this scenario through both theoretical and observational approaches\,\citep[][]{Veske2021ApJ,Liu2021MNRAS,Lu2021MNRAS,Gayathri2020ApJ,Li-Fan2025arXiv}, identifying several potential candidates, such as GW170809-GW151012\,\citep[][]{Veske2020MNRAS} and GW190514-GW190521\,\citep[][]{Li-Fan2025arXiv2}. \citet{Samsing2019MNRASb} estimated the probability of observing two GW merger events originating from a single BSI process. They found that, in the AGN disk scenario, the probability of the second GW merger occurring within ten years of the first is approximately 10\%, whereas in the star cluster case, this probability is below 1\%. Thus, the chance of detecting both GW mergers is only significant for nearly co-planar interactions, which indicates that such double GW merger events could serve as compelling evidence for BSI occurring within AGN disks.


\subsection{Limitations} \label{sec:diss_lim}
In this section, we discuss several limitations of the simulations presented in this paper. \\

$\bullet$ \emph{Resolution limitations}---As illustrated in Figure\,\ref{fig:RunI_distance}, during BSI, the pericenter distances between sBHs can become smaller than the gravitational softening length $\epsilon$, and even smaller than the minimum cell size $\delta_{\rm min}$. Under such conditions, the gas distribution between the BBH components becomes numerically unresolved. We then conduct a higher-resolution simulation for comparison and found that the dynamical evolution of the three-body system does not converge due to the sensitivity of the chaotic three-body system to the gas distribution. Nevertheless, we argue that the BBH merger probability—as a statistical result—remains convergent at the current resolution, for the following reason: The simulation results show that the energy dissipation induced by gas during the first close encounter is consistent between the two resolution levels. This implies that, as long as the orbital parameters—such as pericenter distance and relative velocity—during close encounters between two sBHs are consistent, the resulting energy dissipation due to gas remains consistent. Therefore, although the dynamical evolution of individual chaotic three-body interactions differs between the two resolutions, the statistically averaged results over multiple BSIs process may still converge in our current resolution.

In addition, we set the initial semi-major axis of the BBH to $\sim 0.1\,R_{\rm H}$ and $\sim 0.2\,R_{\rm H}$ with an eccentricity of 0.01 due to resolution limitations. Although newly formed BBHs typically have semi-major axes on the order of $R_{\rm H}/\rm few$ with high eccentricities, these highly eccentric BBHs may undergo decay in both semi-major axis and eccentricity in a dense gaseous environment. For BBHs with smaller initial semi-major axes, the BBH mergers probability during the BSI process is higher in the Gas-Free case\,\citep[see Figure 5 in][]{Fabj2024MNRAS}. However, for three-body systems with higher initial binding energy, we speculate that the gas will have less influence on the BSI process.


\emph{$\bullet$ Accretion of sBH}---Accretion onto sBHs can significantly affect the structure and dynamics of their CSDs, thereby affecting the energy dissipation of the triple system. However, the detailed physics of sBH accretion within AGN disks remains poorly understood. Our simulations suggest hyper-Eddington accretion rates\,($>10^5\,\dot{M}_{\rm Edd}$), which likely induce strong feedback effects, potentially causing substantial deviations from the idealized flows studied here. Moreover, due to resolution limitations, the minimum accretion radius in our simulations is approximately $0.006\,R_{\rm H} > 10^4\,r_{\rm g}$, potentially leading to an overestimation of the accretion rate. In contrast, \citet{Rowan2025arXiv} modeled three non-accreting sBHs. Relative to realistic BSI processes in AGN disks, our accreting setup may underestimate the impact of gas, while the non-accreting model in \citet{Rowan2025arXiv} may overestimate it. This implies that the actual BBH merger probability during BSI in AGN disks could be even higher than reported in this study. Further investigation of this issue is warranted.

 $\bullet$ \emph{Other effects}: More realistic gas equations of state, radiative feedback, outflow in three-dimensional simulation, and gas self-gravity may have an influence on the dynamical evolution during  BSI processes. For instance, radiative feedback and outflow could reduce the accretion rate and change the gas distribution around the sBHs. This may moderately alter energy dissipation during close encounters of BBH. However, due to current technical and computational limitations, these effects are not included in our simulations, which deserves further research in the future. 

\section{Conclusions} \label{sec:summary}
In this work, we study the influence of dense gas in the AGN disk on the BBH merger probability during BSI.
Compared to \citetalias{Wang2025arXiv} and \cite{Rowan2025arXiv}, we perform an updated analysis by incorporating PN2.5 corrections into the $N$-body code and include a significantly larger simulation sample. We conduct a suite of 1800 HD+\emph{N}-body simulations, incorporating PN terms in the \emph{N}-body code to directly obtain the BBH merger probability during BSI. Our key findings are summarized as follows:\\

$\bullet$ \emph{Gas-Induced Enhancements}: The BBH merger probability can be enhanced by a factor of $\sim 5$ in gas-rich AGN disk environments compared to Gas-Free cases, reaching as high as 20–30\%. This enhancement is driven by two gas-induced effects: (i) contraction of the triple system, and (ii) the higher frequency of binary-single encounters during BSI, which sightly dominates the enhancement.

$\bullet$ \emph{Radial Dependence}: The influence of gas on the BBH merger probability during BSI becomes more pronounced with increasing radial distance $R_0$ from the central SMBH. This trend arises because the total gas mass enclosed within the Hill sphere of the three-body system increases as $R_0$ grows. 

$\bullet$ \emph{Comparison with GDF Model}: The GDF approximation fails to replicate the results of HD simulations since it underestimates energy dissipation during close encounters between sBHs.
\\

BBH mergers occurring during BSIs exhibit distinct signatures—such as highly orbital eccentricities, double GW mergers, and GW phase shifts—that set them apart from isolated mergers and motivate targeted searches in future GW detections.

\begin{acknowledgments}
M.W. thanks Yaping Li, Xiangli Lei, Jiancheng Wu, Xiao Fan and Di Wang for helpful discussions. M.W. and Q.W. gratefully acknowledge support from the National Natural Science Foundation of China (grants 123B2041 and 12233007), the National Key Research and Development Program of China (No. 2023YFC2206702) and the National SKA Program of China (2022SKA0120101). Y.M. is supported by the university start-up fund provided by Huazhong University of Science and Technology.  M.W. acknowledge Beijing PARATERA Tech CO., Ltd. for providing HPC resources that have contributed to the research results reported within this paper.
\end{acknowledgments}

\bibliography{ref}{}
\bibliographystyle{aasjournal}
\end{document}